\documentclass{webofc}
\usepackage[varg]{txfonts}   
\usepackage{slashed}
\begin{document}
\title{Special point "trains" in the M-R diagram of hybrid stars}
%
%

\author{
\firstname{David} \lastname{Blaschke}\inst{1}\fnsep\thanks{\email{david.blaschke@uwr.edu.pl}} \and
\firstname{Alexander} \lastname{Ayriyan}\inst{2}\fnsep\thanks{\email{alexander.ayriyan@gmail.com}} \and
\firstname{Mateusz} \lastname{Cierniak}\inst{1}\fnsep\thanks{\email{mateusz.cierniak@uwr.edu.pl}} \and
\firstname{Ana Gabriela} \lastname{Grunfeld}\inst{3,4}\fnsep\thanks{\email{ag.grunfeld@gmail.com}}\and
\firstname{Oleksii} \lastname{Ivanytskyi}\inst{1}\fnsep\thanks{\email{oleksii.ivanytskyi@uwr.edu.pl}}\and
\firstname{Mahboubeh} \lastname{Shahrbaf}\inst{1}\fnsep\thanks{\email{mahboubeh.shahrbafmotlagh@uwr.edu.pl}}
}

\institute{Institute of Theoretical Physics, University of Wroclaw, Max Born Place 9, 50-204 Wroclaw, Poland
\and
IT and Computing Division, A. Alikhanyan National Laboratory, Alikhanian Brothers Str. 2, Yerevan, 0036, Armenia
\and
CONICET, Godoy Cruz 2290, Buenos Aires, Argentina
\and
Departamento de F\'\i sica, Comisi\'on Nacional de Energ\'{\i}a At\'omica, Av. Libertador 8250, (1429) Buenos Aires, Argentina
}

\abstract{%
We present a systematic investigation of the possible locations for the special point (SP), a unique feature of hybrid neutron stars in the mass-radius diagram. The study is performed within the two-phase approach where the high-density (quark matter) phase is described by the covariant nonlocal Nambu--Jona-Lasinio (nlNJL) model equation of state (EOS) which is shown to be equivalent to a constant-sound-speed (CSS) EOS. For the nuclear matter phase around saturation density different relativistic density functional EOSs are used: DD2p00, its excluded-volume modification DD2p40 and the hypernuclear EOS DD2Y-T. In the present contribution we apply the Maxwell construction scheme for the deconfinement transition and demonstrate that a simultaneous variation of the vector and diquark coupling constants results in the occurrence of SP "trains" which are invariant against changing the nuclear matter EOS. 
We propose that the SP train corresponding to a variation of the diquark coupling at constant vector coupling is special since it serves as a lower bound for the line of maximum masses and accessible radii of massive hybrid stars.}
\maketitle

\section{Introduction}
\label{intro}

A key objective of modern NS observations is to infer constraints on the EOS from precise mass and radius measurements. 
The basis for this inference is the unique mapping between the EOS and the stellar structure which is provided by the Tolman-Oppenheimer-Volkoff (TOV) equations.
Recent highlights in this endeavour are the combined mass-radius measurement on the high-mass pulsar PSR J0740+6620 by NICER \cite{Miller:2021qha,Riley:2021pdl}, the multimessenger radius constraint in the $1.4~M_\odot$ mass range \cite{dietrich2020multimessenger} with the tidal deformability constraint from the first binary NS merger GW170817 \cite{abbott2018gw170817}.
While it is clear that for the above mentioned complex of investigations one just needs a function pressure vs. energy density $P(\varepsilon)$ for instance in a multi-polytrope or multi-CSS form as input to the TOV equation in order to solve for the characteristic mass-radius $M(R)$ curve. Doing this a million of times one could straightforwardly extract a boundle of most probable EoS under given observational constraints for $M(R)$ by using Bayesian analysis techniques.
This "agnostic" approach is not satisfactory because it does not allow to infer, e.g., for the composition of matter and at present gives no evidence for a QCD phase transition.

It is clear that between the nucleonic matter phase up to about saturation density $n\lesssim n_0=0.15$ fm$^{-3}$ and the range of validity of perurbative QCD for $n\gtrsim 40~n_0$, there must be a dissociation of hadrons into their quark and gluon constituents.  
Therefore, one should strive to employ a physics educated approach of extracting the most probably EOS, one that inherently includes the QCD phase transition.
For such a type of Bayesian analysis \cite{Ayriyan:2021prr,Shahrbaf:2021cjz}, where the EoS parameters have a direct interpretation as parameters of an effective low-energy QCD Lagrangian, it shall be of eminent importance that their variation will determine the position and character of the deconfinement phase transition. 
As it has been demonstrated by Yudin et al. \cite{Yudin:2014mla} for the case of a simple CSS model of quark matter, varying the bag pressure parameter $B$ determines the onset of deconfinement and leads to the occurrence of a so-called "special point" (SP) in the $M(R)$ 
diagram. The SP is a small region, like a focal point, through which the boundle of all hybrid star sequences with different onset masses $M_{\rm onset}$ for quark deconfinement have to pass. 
Two properties of the SP make it an inevitable ingredient to studies of deconfinement in NS interiors:
1) its existence requires a phase transformation so that it is absent in pure (hadronic or quark) phase stars and can be regarded as a characteristics of hybrid star EOS;
2) its location in the $M-R$ plane is independent of the choice of the hadronic EOS and thus uniquely characterizes the high-density phase of matter.
For a detailed discussion of invariant properties and regions of the SP location in the $M(R)$ diagram for the CSS family of quark matter EOS, see \cite{Cierniak:2020eyh,Blaschke:2020vuy,Cierniak:2021knt,Cierniak:2021vlf}.
A generic Lagrangian model of cold dense quark matter in neutron stars has to have a vector meson interaction channel that provides stiffness and a diquark interaction channel to account for the diquark condensation signalling color superconductivity \cite{Klahn:2006iw,Klahn:2013kga,Baym:2017whm}. 
It came as a surprise when the systematic simultaneous variation of the dimensionless vector meson coupling 
$\eta_V=G_V/G_S$ and diquark coupling $\eta_D=G_D/G_S$ (normalized to the scalar-pseudoscalar coupling strength 
$G_S$) within recent Bayesian analyses studies \cite{Ayriyan:2021prr,Shahrbaf:2021cjz} revealed the existence of series of special points that were lined up in "trains". This phenomenon and its possible consequences for NS phenomenology were discussed first in \cite{Shahrbaf:2021cjz}, but deserve further detailed study as will be outlined in this contribution.

\section{"Trains" of special points}
\label{sec:sp}
The present study is a systematic investigation of the possible locations for SPs, a unique feature of hybrid NS in the mass-radius diagram. It is performed within the two-phase approach where the high-density (quark matter) phase is described in the main part by the covariant nonlocal Nambu--Jona-Lasinio (nlNJL) model EOS which was shown to be equivalent to a CSS EOS \cite{Shahrbaf:2021cjz}. 
The nuclear matter phase up to saturation density and slightly beyond it is described by the relativistic density functional EOS DD2p00 and its excluded-volume modification DD2p40. 
In section \ref{sec:rdf} we discuss the existence of trains of SPs for the case of the confining relativistic density functional approach to color superconducting quark matter and the hypernuclear EoS DD2Y-T \cite{Ivanytskyi:2022bjc,Ivanytskyi:2022oxv} in order to check the robustness of the statements made in the preceeding sections.

\begin{figure}[t]
    \includegraphics[width=0.45\textwidth]{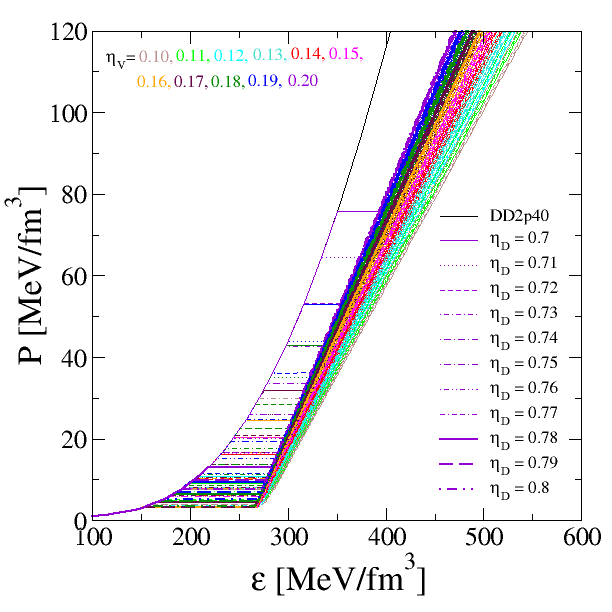}\hfill
    \includegraphics[width=0.45\textwidth]{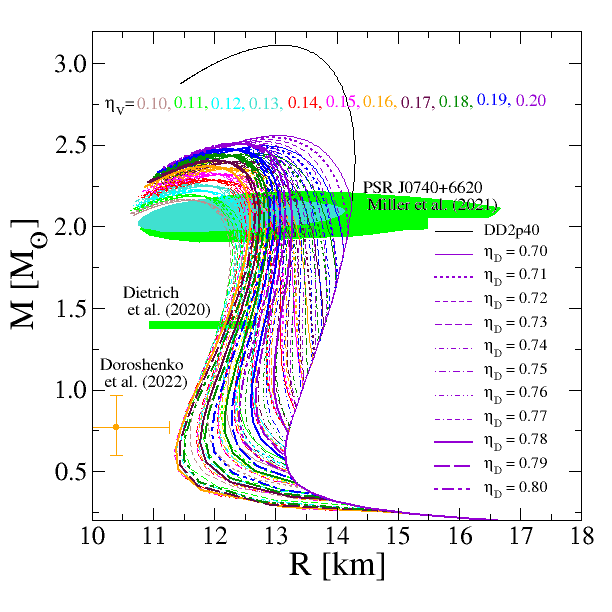}
    \caption{Left panel: family of quark-hadron hybrid EoS obtained within the two-phase approach by a Maxwell construction between the DD2p40 nuclear EOS and the nlNJL quark matter one in its CSS representation.
    Right panel: mass-radius relation for all hybrid star sequences for which the different cases of $\eta_V$ are highlighted in colors and each value of $\eta_D$ corresponds to a line style. Observational constraints on masses and radii are shown, for details see text.}
    \label{fig:M-R_constraint}
\end{figure}

In the left panel of Fig. \ref{fig:M-R_constraint} 
we show the family of hybrid EOS that is obtained by a Maxwell construction with the DD2p40 hadronic EoS and the nlNJL model with the vector meson coupling $\eta_V$ and the diquark coupling $\eta_D$ as two free parameters which are varied in discrete steps.
Details of the model can be found in Ref.~\cite{Shahrbaf:2021cjz}.
As it has been shown in \cite{Shahrbaf:2021cjz}, it is an excellent approximation to employ a CSS form of EOS instead of the numerical solution of the nlNJL model. In that work also a fit formula has been parametrized that maps the pair of free parameters $(\eta_V,\eta_D)$ onto the those of the CSS EOS
\begin{equation}
    \label{eq:CSS_EoS}
    P(\mu) = A \left( 
    {\mu}/{\mu_x}\right)^{1+\beta} - B,
\end{equation}
where $\mu_x = 1$ GeV defines the scale for chemical potential, $A$ is a slope parameter in the units of the pressure, $B$ is the bag pressure and $\beta = 1/c_s^2$ is a parameter related to the squared speed of sound $c_s^2 = dP/d\varepsilon$.

On the right panel of Fig. \ref{fig:M-R_constraint}, the solutions of the TOV equations 
for NS mass versus radius are shown for the EOS depicted on the left panel.
One can see clearly that M-R curves corresponding to a fixed value of $\eta_V$ (shown in the same color) but different values of $\eta_D$ (indicated by different line styles) get collimated in a narrow region close to the maximum mass which has been dubbed SP \cite{Yudin:2014mla}. 

\begin{figure}[ht]
    \includegraphics[width=\textwidth]{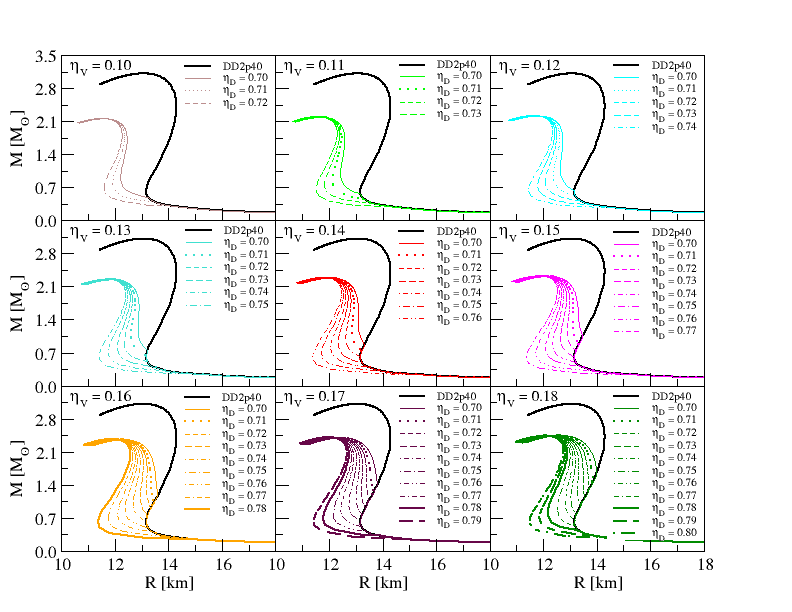}
    \caption{Mass-radius relation for all hybrid stars obtained by a Maxwell construction. The value of $\eta_V$ is taken to be fixed for each panel while the value of $\eta_D$ is varied.}
    \label{fig:M-R_matrix}
\end{figure}

Incrementing the value of $\eta_V$, a new SP is obtained, so that for our set of EOSs a train of SPs in the $M-R$ diagram emerges.
For better visibility of the effect, we have selected a subset of 9 values for $\eta_V$ and show a 3x3 matrix of panels with the corresponding $M-R$ diagrams in Fig.~\ref{fig:M-R_matrix}.

\begin{figure}[ht]
    \includegraphics[width=0.47\textwidth]{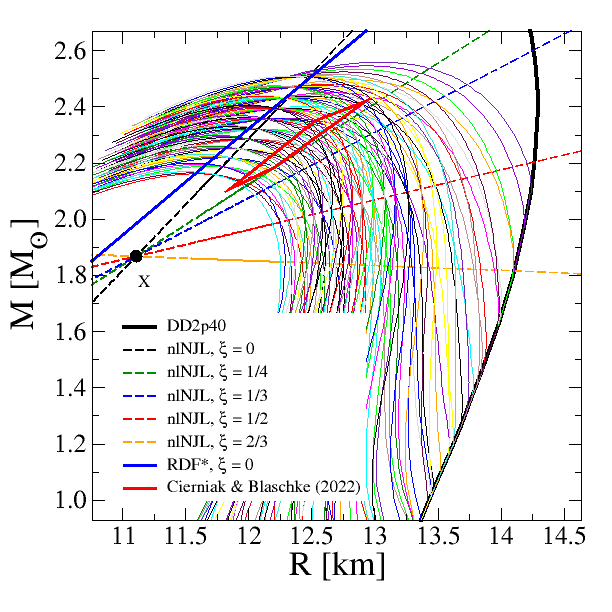}
    \hfill
    \includegraphics[width=0.47\textwidth]{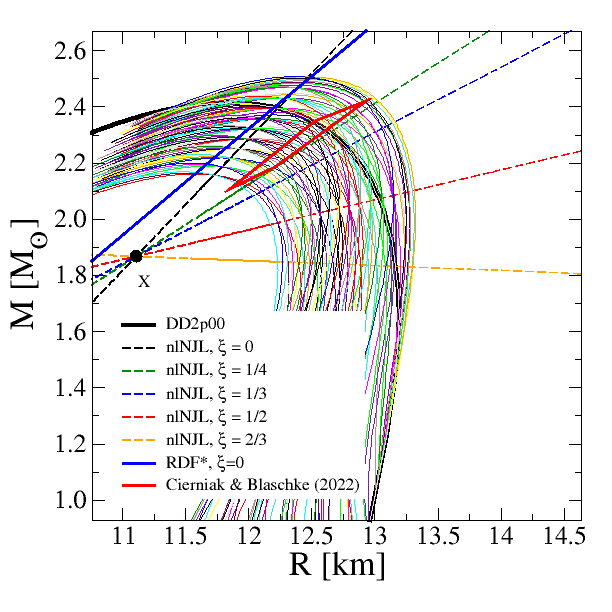}
    \caption{Mass-radius relation for all hybrid stars obtained by a Maxwell construction. Dashed lines connect special points with a fixed slope $\xi=\delta \eta_V/(\delta \eta_D)$ of simultaneous variation of the Lagrangian parameters and meet in one point denoted as X. The lines remain unchanged when the hadronic EOS is changed from the stiff DD2p40 EOS (left panel) to the softer DD2p00 EOS (right panel).
    For a comparison, we show the red trapezoidal region that is enclosed by the upper and lower limits of the parameter $A$ (long edges) and the upper and lower limits of the constant speed of sound (short edges) of the CSS fit to the nlNJL EOS \cite{Cierniak:2021vlf}.
    The bold blue line is obtained from the confining RDF model with conformal limit \cite{Ivanytskyi:2022bjc} for $\xi=0$, see section \ref{sec:rdf}. }
    \label{fig:M-R_lines}
\end{figure}

In Fig. \ref{fig:M-R_lines} we provide a detailed inspection of the location of the special points in the M-R diagram.
We find the following systematics which would hold for a simultaneous variation of the Lagrangian parameters of the quark matter EOS:
\begin{enumerate}
    \item Varying $\eta_V$ and $\eta_D$ simultaneously while keeping the ratio of variations fixed to $\xi=\delta \eta_V/\delta \eta_D$ defines a line $M^{(\xi)}(R)=a_\xi R + b_\xi $ in the $M-R$ diagram along which SPs are located. This line is called a SP train (SPT).
    \item All these SPTs meet in one point denoted by "X" with the coordinates $(M_X,R_X)=(1.8663~\textmd{M}_\odot, 11.112~{\rm km})$.
    \item The slope of the SPTs follows a linear dependence on the parameter $\xi$ 
    $a_\xi = \tan \phi(\xi) = \alpha - \beta \xi$ with $\alpha=0.47074$ and $\beta=0.7252$, see appendix B of Ref.~\cite{Shahrbaf:2021cjz}.
    \item The SPT with largest slope corresponds to $\xi=0$, where each SP on that train corresponds to a fixed $\eta_V$ so that $\delta \eta_V=0$.  This SPT is a lower bound to the line of maximum mass configurations for hybrid stars as a function of $\eta_V$, i.e. the stiffness of quark matter EOS.
    \item We have repeated the analysis by replacing the stiff DD2p40 with the softer DD2p00 EOS. The result is shown in the right panel of Fig. \ref{fig:M-R_lines} and demonstrates that the SPTs are independent of the hadronic EOS used in constructing the hybrid EOS model.  
\end{enumerate}

\begin{figure}[t]
\centering
\includegraphics[width=0.45\columnwidth]{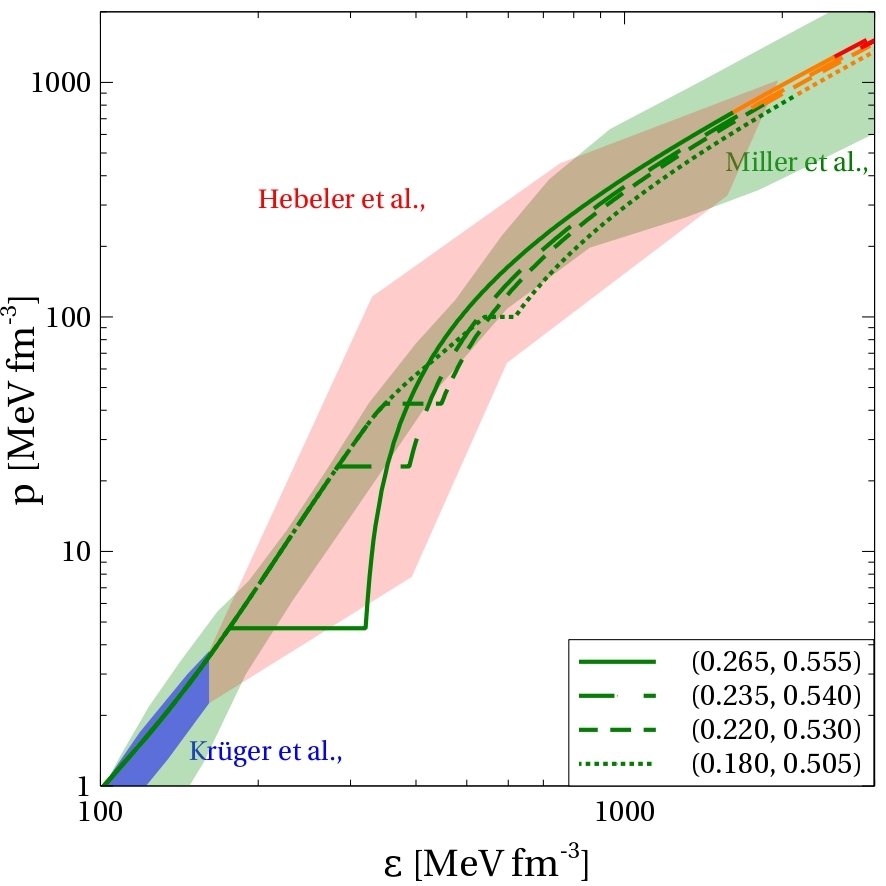}\hfill
\includegraphics[width=0.45\columnwidth]{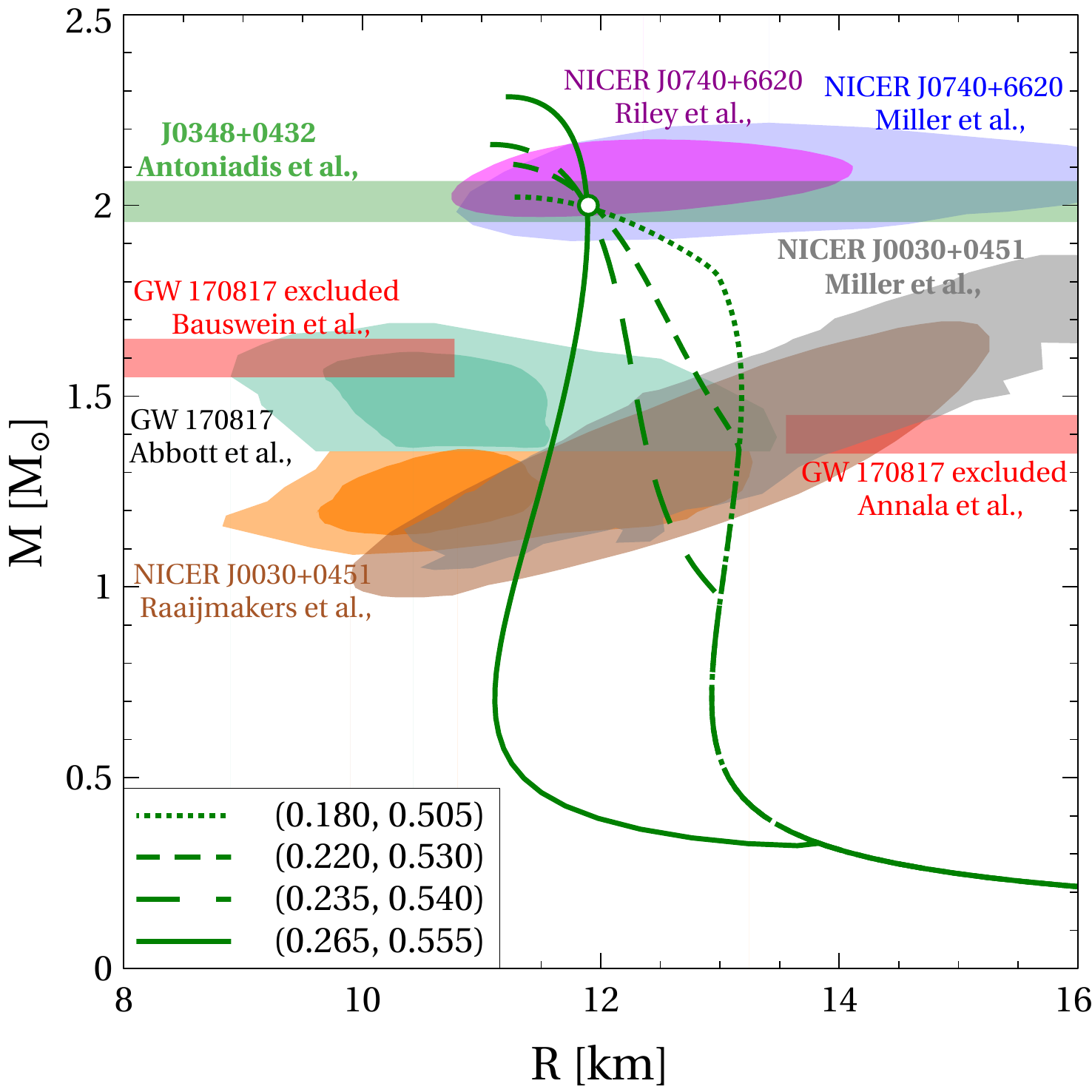}
\caption{The same as in Fig.~\ref{fig:M-R_constraint} but for the confining RDF approach and the hypernuclear EoS DD2Y-T replacing the nlNJL quark matter EoS and the nonstrange hadronic EoS DD2p00 and DD2p40, resp. The pair of values given in brackets in the legends stands for $(\eta_V,\eta_D)$. Figure taken from Ivanytskyi \& Blaschke in these Proceedings.}
\label{fig4}
\end{figure}

\begin{figure}[ht]
   \includegraphics[width=0.65\textwidth]{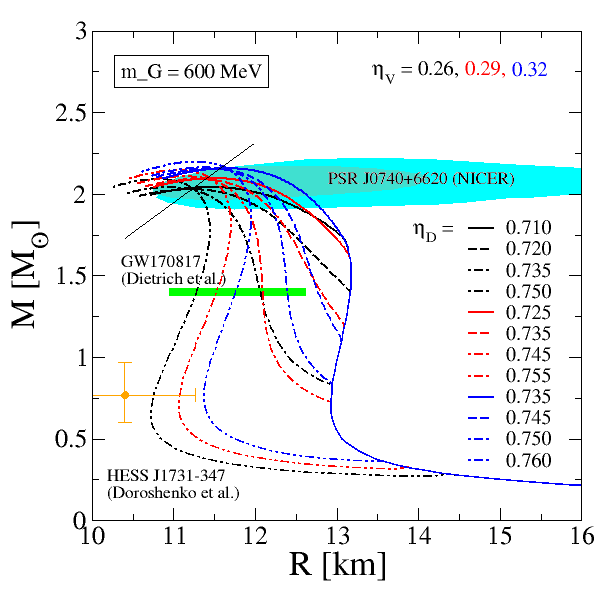}
    \caption{Same as the right panel of Fig.~\ref{fig:M-R_constraint} but for DD2Y-T instead of DD2p40 and the confining RDF approach replacing the nlNJL quark matter EOS. A smaller sample of $\eta_V$ and $\eta_D$ values is used to demonstrate the existence and location of the SP train for the slope $\xi=\delta \eta_V/(\delta \eta_D)=0$ of simultaneous variation of the Lagrangian parameters.}
    \label{fig:MR-rdf}
\end{figure}

\section{Invariance of SPTs under change of the hadronic EOS}
\label{sec:inv}

An important property of the SPs found by varying the bag pressure parameter in the CSS form of the quark matter EOS 
was its independence of the hadronic EOS that was used to obtain the hybrid EOS by a Maxwell construction. This property was proven analytically by Yudin et al. \cite{Yudin:2014mla} and was demonstrated with several examples numerically in \cite{Cierniak:2020eyh}.
Nevertheless, when in the systematic variation of the Lagrangian parameters $\eta_V$ and $\eta_D$ in \cite{Shahrbaf:2021cjz} the appearance of SPs along straight line trajectories came as a surprise, it was not obvious whether these SPTs would remain invariant under changes of the hadronic EOS. 
To give an example of SP non-invariance, in Ref.~\cite{Sen:2022lig} a notable spread of SPs in the mass-radius diagram was obtained for six different hadronic EOS.
The likely reason for this should be the usage of a density-dependent bag pressure functional in the bag model description of the quark matter phase.

It was, however, a very rewarding experience when the invariance of the SPTs could be demonstrated in the investigation of Shahrbaf et al. \cite{Shahrbaf:2021cjz}
for the quark matter phase described by the CSS fit of the color superconducting nlNJL model.
With this invariance and the discovery of the focal point 
$X$ for the SPTs, we have uncovered now a rich phenomenological toolbox for the interpretation of mass-radius measurements in terms of the unknown parameters of Lagrangian models of color superconducting quark matter in the nonperturbative region of low-energy QCD.  
The special role of the $\xi=0$ SPT among all invariant SPTs for the NS phenomenology is explained in the next section. 

\section{Special role of the SPT with $\xi=0$}
\label{sec:xi0}

A closer inspection of the SPTs for $\xi=0$ in the $M-R$ diagrams of Figs. \ref{fig:M-R_matrix}, \ref{fig:M-R_lines} and \ref{fig:MR-rdf} shows that the points where the maximum mass is attained by varying $\eta_D$ at fixed $\eta_V$ lie on a parabola-shaped curve for which the minimum (apex) coincides with the SP. When the maximum mass point (MMP) is to the right of the SP, then the latter lies on the unstable branch and is not interesting. When the MMP is to the left of the SP then the latter is below the maximum mass. Therefore, the SPT for $\xi=0$ can be regarded as a lower bound to the line of highest MMPs for hybrid stars!

Another interesting observation is that for those values of $\eta_D$ for which the SP coincides with the MMP, the onset masses for quark deconfinement are very close to each other.
Therefore, we can identify a particular onset mass $M_{\rm onset}^*$ for which this situation occurs and which is characteristic for the class of quark matter EOS and the hadronic one that has been used to construct the deconfinement phase transition. 
While the SP position does not depend on the choice of the hadronic EOS (as we have seen before), the onset of deconfinement does.
We come to postulate a refined dependence of the maximum mass on the onset mass for deconfinement as
\begin{equation}
\label{eq:Mmax}
M_{\rm max} = M_{\rm SP} + \delta \left|M_{\rm onset}^*-M_{\rm onset} \right|^\kappa~,
\end{equation}
where $\delta$ is a positive constant and $\kappa=2$.
For the nlNJL moodel with a hadronic phase described by DD2p40 we find $M_{\rm onset}^*=0.75~M_\odot$.
A precise definition should be investigated as the closeup view on the SP reveals that it is rather a special region than a point.

\section{SPTs for other quark matter EOS?}
\label{sec:rdf}
A systematic study with different classes of Lagrangian quark matter models is under way. All models shall have in common that they have vector and diquark coupling strengths that are not determined by vacuum observables so that they are considered as free parameters like in the recent works in this direction, e.g.,  \cite{Baym:2017whm,Ayriyan:2021prr,Shahrbaf:2021cjz,Ivanytskyi:2022oxv}.  
Currently, the instantaneous nlNJL model with 3-momentum dependent nonlocality formfactor \cite{Contrera:2022tqh} is under investigation.

In this contribution, we can report first results that are obtained within the confining RDF approach to color superconducting quark matter \cite{Ivanytskyi:2022bjc}, where the existence of SPs was observed in \cite{Ivanytskyi:2022bjc,Ivanytskyi:2022oxv} for the regime when pairs of couplings $(\eta_V,\eta_D)$ were chosen along a straight line in the $\eta_D-\eta_V$ plane, see Fig.~\ref{fig4}. 
The analogue of a $\xi=0$ line was found (see Fig.~\ref{fig:MR-rdf}) and it compares well with the one from the covariant nlNJL model, see the bold blue line in both panels of Fig.~\ref{fig:M-R_lines}.
Also in this case the relation (\ref{eq:Mmax}) between $M_{\rm max}$ and $M_{\rm onset}$ holds, but with 
$M_{\rm onset}^*=1.4~M_\odot$ being larger than for the nlNJL case with DD2p40.

It is interesting to note that the confining RF approach with an early onset of deconfinement leads to a dramatic compactification of the light hybrid star for masses below $2~M_\odot$ in agreement with all present radius constraints in three mass ranges that follow for the strangely light central compact  object in HESS J1731-347 \cite{doroshenko2022strangely}, the multimessenger radius constraint at $1.4~M_\odot$ \cite{dietrich2020multimessenger} and radius measurement on the high-mass pulsar J0740+6620 by NICER \cite{Miller:2021qha,Riley:2021pdl}. Comparing with Fig.~\ref{fig:M-R_constraint} one observes that the early deconfinement hybrid star sequences for the nlNJL model of color superconducting quark matter cannot describe the radius of $R=10.4^{+0.86}_{-0.78}$ km at $M=0.77^{+0.20}_{-0.17}~M_\odot$ reported by Doroshenko et al. in \cite{doroshenko2022strangely}.

What about the original CSS model for quark matter, which was used in the original work by Yudin et al. \cite{Yudin:2014mla} who started the investigation of the SP
property? 
Their analytic investigation used a two-parameter bag model EoS for which the speed of sound was fixed to $c_s^2=1/3$ and varying the bag constant $B$ produced a varying onset of deconfinement and defined the SP. It was shown analytically that the SP position depends only on the EOOS of the high-density phase, not of the hadronic one. A variation  of the speed of sound parameter was not performed in \cite{Yudin:2014mla}. 
The next systematic investigation of the SP property of hybrid EOS was performed in \cite{Cierniak:2020eyh} with a three-parameter CSS EoS (\ref{eq:CSS_EoS}) for quark matter. 
The invariance of the SP against variation of the hadronic EOS was confirmed and in follow-up works \cite{Cierniak:2021knt,Cierniak:2021vlf} the location of the SP (still defined by varying $B$) in dependence of the values for the remaining two parameters $c_s^2$ and $A$ was explored. In a sense, also in these studies trajectories 
in the mass-radius diagram were found upon variation of the parameter pair $(c_s^2,A)$. The accessible region is shown in Fig.~\ref{fig:M-R_lines} as a bold red trapezoid with borders defined by the limiting values for $c_s^2$ and $A$ that follow from the CSS fit to the nlNJL quark matter EOS. 
The discrepancy to the $\xi=0$ SPT obtained from the nlNJL model (black dashed line) or the confining RDF model (bold blue line) is apparent. Obviously it makes a difference how the SP position is defined. It would be interesting to explore the trajectory of SPs for the CSS model when the bag constant is kept fixed and only the parameter pair $(c_s^2,A)$ is varied. 

\section{Conclusions}
\label{concl}
In the present contribution we have applied the Maxwell construction scheme for the deconfinement phase transition and demonstrated that a simultaneous variation of the vector and diquark coupling constants results in the occurrence of SP "trains" which are invariant against changing the nuclear matter EOS. We have shown that the phenomenon occurs also in more sophisticated Lagrangian quark matter models such as the confining relativistic density functional approach and should be generic to all classes of multiparameter Lagrangian models.
We have pointed out that Bayesian approaches appropriate for testing the presence of a deconfinement phase transition in NSs should generate SPTs upon parameter variation.
We have proposed that the SPT corresponding to a variation of the diquark coupling at constant vector coupling is special since it serves as a lower bound to the line of maximum masses and accessible radii of massive hybrid stars.

\section*{Acknowledgements}
We are grateful to K. Redlich for his interest in our work and stimulating discussions. 
This work was supported by NCN, grant 2019/33/B/ST9/03059 (A.A., D.B., M.C., O.I., M.S.) and Consejo Nacional de Investigaciones Cient\'ificas y T\'ecnicas, grant~PIP17-700 (A.G.G).



\end{document}